# On The Method of Precise Calculations Of Energy Levels of Hydrogen-like Atoms

## N. A. Boikova, Y. N. Tyukhtyaev, R. N. Faustov

**Abstract:** *We describe a method for deriving logarithmic corrections in the mass ratio to the S-level of a hydrogen-like atom. With this method, a number of new corrections of this type are calculated analitically for the first time.*

The comparison of theoretical and experimental results on non-relativistic bound states in QED is an important source of information about the fundamental physical parameters, such as fine structure constant, the masses of electron, muon, proton and many others. The promotion of theoretical result precision requires more and more difficult calculations of high order corrections. The solution of this problem is impossible without the development of bound state theory and should use the new methods of calculation.

Long debates on the logarithmic contribution ($\alpha^6 \ln \alpha$) to the fine shift of *S*-levels in muonium took place in resent years. It was found eventually that such correction vanishes. However, it is highly probable that there is correction involving the logarithm of parameter $\beta = m_1 / m_2$.

In the review [*1*] the published earlier recoil corrections are summarized. This table looks as follows: (*Table 1*). Among the recoil corrections there is only one logarithmic contribution in the parameter $\beta$ [*2*]. The problem of existence of such different corrections is discussed on the basis of the quasipotential approach in [3-6].

Table 1. Recoil Correction to Lamb Shift

|  | $\dfrac{(Z\alpha)^5}{\pi n^3}\dfrac{m_r^3}{mM}$ | $\Delta E(1S)$ kHz | $\Delta E(1S)$ kHz |
|---|---|---|---|
| *Coulomb-Coulomb Term* <br> Salpeter (1952) <br> Fulton, Martin (1954) | $-\dfrac{4}{3}\delta_{10}$ | -590.03 | -73.75 |
| *Transverse-Transverse Term* <br> *Bulk Contribution* <br> Salpeter (1952) <br> Fulton, Martin (1954) <br> Erickson, Yennie (1965) <br> Grotch, Yennie (1969) <br> Erickson (1977) <br> Erikcson, Grotch (1988) | $\{2\ln\dfrac{2Z\alpha}{n} + 2[\psi(n+1)-\psi(1)]$ <br> $+\dfrac{n-1}{n} + \dfrac{8(1-\ln 2)}{3}\}\delta_{10}$ <br> $-\dfrac{1-\delta_{10}}{l(l+1)(2l+1)}$ | -2 494.01 | -305.46 |
| *Transverse -Transverse Term* <br> *Very high momentum Contribution* <br> Fulton, Martin (1954) <br> Erickson (1977) | $-\dfrac{2}{M^2-m^2}\left(M^2 \ln\dfrac{m}{m_r} - m^2 \ln\dfrac{M}{m_r}\right)\delta_{10}$ | -0.48 | -0.06 |
| *Coulomb-Transverse Term* <br> Salpeter (1952) <br> Fulton, Martin (1954) <br> Erickson, Yennie (1965) | $\tfrac{8}{3}\{[\ln\tfrac{2}{nZ\alpha} + [\psi(n+1)-\psi(1)]$ <br> $-\tfrac{1}{2n} + \tfrac{5}{6} + \ln 2]\delta_{10} - \ln k_0(n,l)$ | 5 494.03 | 720.56 |
| Grotch, Yennie (1969)<br>Erickson (1977)<br>Erickson, Grotch (1988) | $-\frac{1-\delta_{l0}}{2l(l+1)(2l+1)}\}$ | | |
| $\Delta E(nS)$<br>Pachucki, Grotch (1993)<br>Eides, Grotch (1997) | $(4\ln 2 - \frac{7}{2})(\pi Z\alpha)\delta_{l0}$ | -7.38 | -0.92 |
| $\Delta E(nL)(l\neq 0)$<br>Golosov, Elkhovski,<br>Milshtein, Kriplovich (1995)<br>Jentschura, Pachucki (1996) | $\frac{3(1-\frac{l(l+1)}{3n^2})(\pi Z\alpha)}{4(l-\frac{1}{2})(l+\frac{1}{2})(l-\frac{3}{2})}$ | | |
| $\Delta E(nS)$<br>Pachucki, Karshenboim (1993)<br>Melnikov, Elkhovski (1999) | $-\frac{11}{15}(Z\alpha)^2\ln^2(Z\alpha)\delta_{l0}$ | -0.42 | -0.05 |

The three-dimensional relativistic approach to the description of bound states of two particles is developed [7-9] and allows one to find various logarithmic corrections in the parameter $\beta$ to the magnitude of the fine shift of S-level energies in hydrogen-like atoms in sixth orders in the fine structure constant and to determine the conditions under which the contribution found in [2] is the only one.

We consider the methods of quasipotential definition, used for the description of bound states of two particles. The equation for the Green function of two fermions can be written in a Schwinger form:

$$\{(\gamma\pi - M)_1(\gamma\pi - M)_1 - I_{12}\}G = I \tag{1}$$

where $\pi = p - eA$; $p$ is the 4-momentum of the particle, $A$ is the external field, $M$ is the mass operator of the particle, $I_{12}$ is the interaction kernel of particles 1,2 and $I$ is the unit operator. Equation (1) can also be written in Bethe-Salpeter form:

$$G = g_0 + g_0 I_{12} G, \quad g_0 = iG_1 G_2 \tag{2}$$

where $G_i$ is the free fermion function. The Coulomb kernel is separated as the main part of interaction.

$$I_{12} = K_C + \tilde{I}, \quad K_C(\vec{p},\vec{q}) = -\frac{e_1 e_2 \gamma_{10}\gamma_{20}}{(\vec{p}-\vec{q})}. \tag{3}$$

Then the Coulomb Green function is determined by the equation

$$\left[(\gamma\pi - M)_1(\gamma\pi - M)_2 - K_C\right]g_C = I,$$
$$g_C = g_0 + g_0 K_C G_C. \tag{4}$$

Using Eq. (1) and (4) the Green's function can be written as

$$G = g_C + g_C \tilde{I} G. \tag{5}$$

To include the mass corrections into the interaction kernel it is convenient to use the expressions

$$G = G_C + G_C \tilde{K} G, \quad G_C = G_0 + G_0 K_C G_C,$$
$$G_0 = i S_1 S_2, \quad S_i^{-1} = (\hat{p} - m)_i.$$
(6)

$$\tilde{K} = I_{12} - \left[ M_1 S_2^{-1} + S_1^{-1} M_2 - M_1 M_2 \right] + K_C = K - K_C.$$
(7)

The energy levels of a composite particle are poles of the Coulomb propagator as a function of $E$. Therefore it is useful to express the full Green's function via the Coulomb Green's function and to expand in a series as

$$G = G_C - G_C (\tilde{K}^{(1)} + \tilde{K}^{(2)} + \tilde{K}^{(1)} G_C \tilde{K}^{(1)} + ...) G_C.$$
(8)

The kernel index shows the order of kernel with respect to interaction constant $\alpha$. The order of the terms of perturbation theory series is determined by the number of transverse photons exchanges. The terms in series (8) are of the same order in $\alpha$. That is why near the poles the expansion of Coulomb Green's function $G_C$ isn't truncated. It is necessary to sum the infinite series of diagrams. Using the following equation

$$G = G_C + G_C \tilde{K} G = G_C + G \tilde{K} G_C,$$
(9)

we can introduce scattering amplitude

$$\tilde{T} = \tilde{K} + \tilde{K} G \tilde{K},$$
(10)

and write the full Green's function as

$$G = G_C + G_C \tilde{T} G_C.$$
(11)

Using the integration over the relative energies and projection on positive-frequency states, we obtain:

$$\widehat{G}^+ = \widehat{G}_C^+ + \overline{(G_C \tilde{T} G_C)}^+.$$
(12)

Then the inverse Green's function is represented as

$$(\widehat{G}^+)^{-1} = (\widehat{G}_C^+)^{-1} - (\widehat{G}_C^+)^{-1} \overline{(G_C \tilde{T} G_C)}^+ (\widehat{G}^+)^{-1}.$$
(13)

The wave function of the two-fermion system with the total energy $E$ satisfies the quasipotential equation

$$\{(\widehat{G}^+(\vec{p}, \vec{q}; E))^{-1} - \tilde{V}(\vec{p}, \vec{q}; E)\} \Psi_E(\vec{q}) = 0,$$
(14)

where the quasipotential is determined as

$$\tilde{V} = (\widehat{G}_C^+)^{-1} \overline{(G_C \tilde{T} G_C)}^+ (\widehat{G}^+)^{-1}.$$
(15)

This definition is the most general expression for two particles interaction quasipotential. The expansion of quasipotential has the form

$$\tilde{V} = \tilde{\tau}_C - \tilde{\tau}_C \widehat{G}_C^+ \tilde{\tau}_C + ..., \quad \tilde{\tau}_C = (\widehat{G}_C^+)^{-1} \widehat{G_C \tilde{T} G_C}^+ (\widehat{G}_C^+)^{-1}. \tag{16}$$

We consider the quasipotential in the lowest approximation

$$\tilde{V}^{(2)} \approx \tilde{\tau}_C. \tag{17}$$

Discuss a most simple case, when the amplitude is defined as $\tilde{T}^{(2)} = K_T$,

$$K_T = -\frac{4\pi\alpha}{k_0^2 - \vec{k}^2 + i\varepsilon}(\vec{\gamma}_1\vec{\gamma}_2 - \frac{(\vec{\gamma}_1\vec{k})(\vec{\gamma}_2\vec{k})}{k^2}), \tag{18}$$

where $K_T$ characterizes the exchange with transverse photon. The bound effects can be considered in the approximation

$$G_C \approx G_0 + G_0 K_C G_0. \tag{19}$$

This approximation allows one to describe the exchange between Coulomb and transverse photons. Using the equation for Green's function of noninteractive particles and the result of projection on positive frequency states the inverse Green's function is represented as

$$(\widehat{G}_C^+)^{-1} = F^{-1} - K_C^+, \quad F = \widehat{G}_0^+. \tag{20}$$

Then the amplitude $\tau_C$ is defined by

$$\tilde{\tau}_C = (K_T)_{0F}^+ + (K_C G_0 K_T)_{0F}^+ - K_C^+ F(K_T)_{0F}^+ + (K_T G_0 K_C)_{0F}^+ - (K_T)_{0F}^+ F K_C^+ + (K^{(2)})_{0F}^+, \tag{21}$$

where symbol $(X)_{0F}^+$ means $F^{-1} \widehat{X}^+ F^{-1}$, $K_{(2)} = K_{CT} + K_{TC}$. Excluding the main Coulomb interaction and considering the exchanges by Coulomb and transverse photons we can write the quasipotential as

$$V(\vec{p}, \vec{q}; E) = F^{-1} - (\widehat{G}^+)^{-1} = F^{-1} - \tau_0 + \tau_0 F \tau_0 + ..., \tag{22}$$

where $\tau_0 = F^{-1} \widehat{G_0 T G_0}^+ F^{-1}$, $T = K + KGK$. \hfill (23)

Then using the Coulomb gauge for photon propagator we obtain

$$\tau_0 \cong (K_T)_{0F}^+ + (K_C G_0 K_T)_{0F}^+ + (K_T G_0 K_C)_{0F}^+ + (K_{CT})_{0F}^+ + (K_{TC})_{0F}^+ \tag{24}$$

Expression (24) is included in (21). Therefore, the block $\tau_0$ is the part of expression $\tau_C$ and quasipotential $\tilde{V}$. Note, that if the quasipotential is determined in term of $\tau_C$, the reducible graphs and corresponding iterations are considered in the first order of perturbation theory. However, using the quasipotential (22) we obtain iterations only in the second order of expansion.

$$V^{(2)} = (K_T)_{0F}^+, \tag{25}$$

$$V^{(4)} = (K_C G_0 K_T)^+_{0F} + (K_T G_0 K_C)^+_{0F} - K_C^+ F(K_T)^+_{0F},$$
$$-(K_T)^+_{0F} F K_C^+ + (K_{CT} + K_{TC})_{0F}.$$
(26)

The presence of iteration terms for each reducible graph, improved its behavior in the infrared region, is the main feature of the quasipotential. It should be noted that, because of the arising difference of reducible graphs and it's iteration terms contributions, the main corrections to the Coulomb energy level from $V^{(2)}$ and $V^{(4)}$ are in one case, of lower order, and, in other case, of higher order in the fine structure constant $\alpha$. As the number of events of Coulomb photon exchange increases in higher orders of perturbation theory, the conditions for the cancellation of the leading corrections in the differences of reducible diagrams and the corresponding iterations remain unchanged. Since the iterations correspond only to reducible graphs, the nonreducible diagram with single transverse photon is considered taking into account the multiple exchanges with Coulomb photons.

The described methods of constructing the quasipotential (15) and (22) are not isolated and constitute the system, connected by consistent transitions from common to the particular case. For example, to describe the interactions of particles in hydrogen-like atoms in the low-frequency domain the quasipotential is used in the form

$$\tilde{V} = \frac{\tau_C}{1 + \widehat{G}_C^+ \tau_C}.$$
(27)

The terms of $G_C$ expansion series are of the same order in $\alpha$, therefore one can use the nonrelativistic analog $G_C$, that is the Green's function of the Schrödinger equation with Coulomb potential.

In the high-frequency domain one can iterate the equation (6) for Coulomb Green's function and obtain the following expression for quasipotential according to (27)

$$V = \frac{\tau_0}{1 + F\tau_0}$$
(28)

Performing the contour integration by zero components of momenta we can simplify the quasipotential (28). Using the Fourier representation of $\delta$ function the block $\overline{G_0 T G_0}$ from $\tau_0$ expression can be written as

$$\overline{G_0 T G_0} = -(2\pi)^{-4} \int_{-\infty}^{\infty} dt \int_{-\infty}^{\infty} d\tau \int_{-\infty}^{\infty} dp_0 e^{ip_0 t} S_1(p_1) S_2(p_2) \int_{-\infty}^{\infty} dk_0 e^{-k_0 t} \otimes$$
$$\otimes \int_{-\infty}^{\infty} dk'_0 e^{ik'_0 \tau} T(k_0, k'_0, \vec{p}, \vec{q}; E) \int_{-\infty}^{\infty} dq_0 e^{iq_0 \tau} S_1(q_1) S_2(q_2).$$
(29)

Taking the fermion propagator expressions via projection operators

$$S_1(E_1+p_0,\vec{p}) = \left(\frac{\Lambda_1^+(\vec{p})}{E_1+p_0-\varepsilon_{1p}+i\varepsilon} + \frac{\Lambda_1^-(\vec{p})}{E_1+p_0-\varepsilon_{1p}-i\varepsilon}\right)\gamma_{10},$$

$$S_2(E_2-p_0,-\vec{p}) = \left(\frac{\Lambda_2^+(-\vec{p})}{E_2-p_0-\varepsilon_{2p}+i\varepsilon} + \frac{\Lambda_2^-(-\vec{p})}{E_2+p_0-\varepsilon_{2p}-i\varepsilon}\right)\gamma_{20}, \quad \varepsilon_{pi}=\sqrt{p^2+m_i^2}.$$

and using residue theory one can obtain the expression for $\tau_0$:

$$\tau_0 = T_+(\vec{p},\vec{q};E) + \sum_{i=1}^{2}(\varepsilon_{ip}-E_i)\Delta T'_{i+}(\vec{p},\vec{q};E) + \sum_{i=1}^{2}(\varepsilon_{iq}-E_i)\Delta T''_{i+}(\vec{p},\vec{q};E) +$$
$$+\sum_{i,k=1}^{3}(\varepsilon_{ip}-E_i)\Delta T'''_{ik+}(\vec{p},\vec{q};E)(\varepsilon_{kq}-E_k).\qquad(30)$$

In many cases for the investigation of energy levels of hydrogen-like atoms up to the accuracy of terms of the order $\alpha^5$ we can restrict ourselves by the first term of the sum in expression (30) that is we can use

$$\tau_0 \approx T_+(\vec{p},\vec{q};E) \;,\; V = \frac{T_+}{1+FT_+}.\qquad(31)$$

Because in the weakly coupled systems the particles located near the mass shell one can use the scattering approximation to consider two photon interactions up to the accuracy of terms of the order $\alpha^5$ and replace $T_+(\vec{p},\vec{q};E)$ with $T_+(0,0;E)$

$$T_+(\vec{p},\vec{q};E) \to T_+(0,0;E).\qquad(32)$$

In the high virtual momenta domain the matrix elements of scattering amplitude can be approximated on the mass shell and calculated at values of the momenta $|\vec{p}|\sim|\vec{q}|\cong 0$, $E\cong m_1+m_2$. The on-shell scattering amplitude has poles in the infrared region. The infrared divergences are canceled by the introduction of some minimum virtual momenta $k_{min}\cong\varepsilon$. The concrete value $\varepsilon$ is not important, because the infrared divergences are canceled in the sum of the diagrams. Then one can perform the change

$$T_+(0,0;E) \to T_+^0(0,0,m_1+m_2),\qquad(33)$$

and the quasipotential is determined as

$$V = \frac{T_+^0(0,0,m_1+m_2)}{1+FT_+^0(0,0,m_1+m_2)}.\qquad(34)$$

We note that this quasipotential does not allow one to calculate exactly the corrections up to the accuracy $\alpha^6\ln\alpha$ to the shift of $S$ energy levels of hydrogen-like atoms. In our papers [8,10] it is shown that for more full calculation of contributions order $\alpha^6 \ln \alpha^{-1}$ to the hyperfine shift of muonium it is required to use quasipotential, expressed via the amplitude $\tau_0$ (28).

Thus, to consider the concrete problem of bound states different methods of constructing the quasipotential are used. The connection among them is expressed in the possibility of approximations use for amplitude and can be describe schematically:

$$\tau_c \to \tau_0 \to T_+ \to T_+(0,0;E) \to T_+^0(0,0,m_1+m_2). \tag{35}$$

Let us employ the method using Coulomb Green function to analyze the hyperfine shift in positronium and to find out limitations of the scheme constituents.

The information about the corrections to the Coulomb energy levels can be obtained by constructing perturbation theory on the basis of the above arguments and be solved quasipotential equation:

$$(E-\varepsilon_{1p}-\varepsilon_{2p}) \Psi_E(\vec{q}) = (2\pi)^{-3} \int d^3q \, \tilde{V}(\vec{p},\vec{q};E) \, \Psi_E(\vec{q}) \tag{36}$$

where $\Psi_E(\vec{q})$ is the wave function of the two-fermion system with total energy $E$. Solving this equation, we find approximately

$$\psi_E(\vec{p}) = \Omega_p \varphi_C(\vec{p}), \quad \Omega_p = \frac{M_{p1}M_{p2}}{2\mu(M_{p1}+M_{p2})},$$

$$\varphi_C(\vec{p}) = 8\pi\alpha\mu\varphi_C(0)\varphi_p^2; \quad \varphi_C(0) = \frac{(\alpha\mu)^{3/2}}{\sqrt{\pi}}; \quad \varphi_p = (p^2+\alpha^2\mu^2)^{-1} \tag{37}$$

where $M_{pi}=\varepsilon_{pi}+m_i, \varepsilon_{pi} = \sqrt{p^2+m_i^2}$, $\varphi_C(\vec{p})$ - is a solution of the nonrelativistic Schrödinger equation with the Coulomb potential. In accordance with the perturbation theory the second order corrections are given by

$$\Delta E_n = \langle \psi_{ns} | \, \Delta V^{(2)} + V^{(4)} + \sum_{m \neq n} \{ \Delta V^{(2)} |\psi_{ms}\rangle \frac{1}{E_n - E_m} \langle \psi_{ms} | \, \Delta V^{(2)} \} |\psi_{ns}\rangle. \tag{38}$$

It is clear from the form of the Coulomb wave function that the main contribution to the splitting of the energy levels comes from momenta $p^2 \sim Z^2\mu^2\alpha^2$. That is why it is possible to use the expansion of the quasipotential in $p^2/m_i^2$. In accordance with (38) the contribution of hyperfine shift of the S energy levels in the positronium is defined in the lowest order

$$\Delta E = (2\pi)^{-3} \int \varphi_C^*(\vec{p}) \, \Delta V^{(2)}(\vec{p},\vec{q}) \, \varphi_C(\vec{q}) \, d^3p \, d^3q, \tag{39}$$

After the expansion in $p^2/m_i^2$ we obtain for the diagram of one photon exchange

$$V^{(2)}(\vec{p},\vec{q}) = \frac{e^2}{|\vec{p}-\vec{q}|^2}(1+\frac{1}{4m^2}((\vec{\sigma}_1(\vec{p}-\vec{q}))(\vec{\sigma}_2(\vec{p}-\vec{q}))-(\vec{p}-\vec{q})^2(\vec{\sigma}_1\vec{\sigma}_2))) \quad (40)$$

After the integration expression (39) we find the main contribution of the order $\alpha^4$:

$$\Delta E = \frac{2\pi\alpha}{3m^2}|\varphi_C(0)|^2\langle\vec{\sigma}_1\vec{\sigma}_2\rangle. \quad (41)$$

However, it is possible to separate this contribution using the approximation:

$$\varphi_C(\vec{p}) = (2\pi)^{3/2}\varphi_C(0)\delta(\vec{p}), \quad E = m_1 + m_2. \quad (42)$$

If we restrict ourselves to the accuracy $O(\alpha^5)$, then to investigate the contributions of two photon diagrams one can also use this approximation, since $T^{(4)} \sim \alpha^2$ and $|\varphi_c(0)|^2 \sim \alpha^3$. In this case, to exclude infrared divergences we cut the virtual momentum of the low-frequency region. In [11-13] it is shown that the summary contribution of two photon diagrams is final and doesn't depend on the cutoff factor. So, the hyperfine splitting of the ground state up to the accuracy of $\alpha^5$ is calculated using the approximation (42) in the simplest way.

It is shown [14] that the contribution of three-photon exchanges includes the uncancelled cutoff factor of the virtual momentum. Because of this it is necessary to use the Coulomb wave function and to take into account the quasipotential dependence on the momenta and full energy.

The retardation effects are considered more accurately when the amplitude $T_+(\vec{p},\vec{q};E)$ is used. The corresponding method of constructing the quasipotential allows one to obtain the correction of order $\alpha^6\ln\alpha$. However, to calculate the logarithmic contributions completely it is necessary to use the amplitude $\tau_0$ and the relativistic modification of the Coulomb wave function [15].

So, according to our investigations [10,16-18] the result for the hyperfine splitting of the ground energy levels in the positronium atoms was established up to the accuracy of $\alpha^6\ln\alpha$.

$$\Delta E_{hfs} = \frac{\alpha^4 m}{2}\left[\frac{7}{6} - \frac{\alpha}{\pi}(\frac{16}{9}+\ln 2) + \frac{5\alpha^2}{12}\ln\alpha^{-1}\right]. \quad (43)$$

The result (43) was obtained also in [19,20].

Once the theoretical result for the hyperfine splitting of the ground energy levels in the muonium and positronium atoms had been reliably established up to the accuracy of $\alpha^6\ln\alpha$ effort to study the effect of recoil of the heavy particles was intensified. According to [15] the relativistic factor can give rise to logarithmic corrections only in the lowest order in $\alpha$. Taking this into account we analyzed the effect of recoil on the fine shift of the S levels of the muonium atom from the one photon exchange first.

The analysis of one photon interaction in the muonium atom contains the following parts. At first using residue theory the contour integration is performed. Upon this integration the integrands factorize and the fine shift takes a three-dimensional form. Then using the symmetries of the expressions the matrix structure is transformed. After the simplification of the matrix structure, the required estimates lead to integrals making contributions to the fine shift that are logarithmic in the ratio of the particle masses.

$$i_{ST} = \frac{\alpha^6 \mu^3}{m_1 m_2} \beta \int \frac{d\vec{q}}{\sqrt{q^2+\beta^2}(q^2+\gamma^2)} \int \frac{d\vec{p}}{\sqrt{p^2+\beta^2}(\vec{p}+\vec{q})^2\sqrt{p^2+1}}. \qquad (44)$$

$$j_{ST} = \frac{\alpha^6 \mu^3}{m_1 m_2} \beta \int \frac{d\vec{q}}{\sqrt{q^2+\beta^2}\sqrt{q^2+1}(q^2+\gamma^2)} \int \frac{d\vec{p}}{\sqrt{p^2+\beta^2}\sqrt{p^2+1}(q^2+\gamma^2)}. \qquad (45)$$

It is established for the hydrogen-like atom involving a nucleus of charge $Ze$, that the general expression for the shift of the energy levels due to one photon interaction is given by

$$\delta E_T(n,l) = \frac{(Z\alpha)^6 \mu^5}{m_1^2 m_2^2 n^3} \frac{2\sqrt{2}}{\pi^2} \left\{ \ln\left(1+\sqrt{2}\right) - \sqrt{2} - \beta \ln \beta^{-1} \right\} \ln \beta^{-1}. \qquad (46)$$

By estimating the frequencies corresponding to these shifts of the energy levels we arrive at

$\delta \nu_m(1S) \approx 3{,}8$ kHz

$\delta \nu_m(2S) \approx 0{,}48$ kHz  $\qquad\qquad\qquad\qquad\qquad\qquad\qquad\qquad\qquad (47)$

$\delta \nu_m(2S-1S) \approx -3{,}32$ kHz

Note for comparison that the logarithmic contribution $\alpha^6 \ln \alpha$ to the hyperfine shift of the ground energy levels in the muonium atom is equal to 11,29 kHz.

In analyzing two-photon diagrams, it turns out that one must take into account the multiple exchanges of Coulomb photons. However, the effects of binding manifest themselves only in the low-frequency region at virtual momentum values of $|\vec{k}| < \varepsilon$, $\mu(Z\alpha)^2 \leq \varepsilon \leq \mu$, since the binding energy of a hydrogen-like atoms is small. In view of this it is reasonable to break down the interval of integration into the low- and the high-frequency regions.

In the low-frequency domain the nonrelativistic approximation can be efficiently applied. The quasipotential is defined via the amplitude $\tau_c$ and is given by

$$\tilde{V} = \tau_c - \tau_c \widehat{G}_c^+ \tau_c + ... \approx T_+ - T_+ g_c T_+. \qquad (48)$$

In the lowest approximation

$T = K^{(1)} + K^{(2)}$ , $K^{(1)} = K_T$. $\qquad\qquad\qquad\qquad\qquad\qquad\qquad\qquad (49)$

The kernel $K^{(2)}$ is shown *Fig.1*.

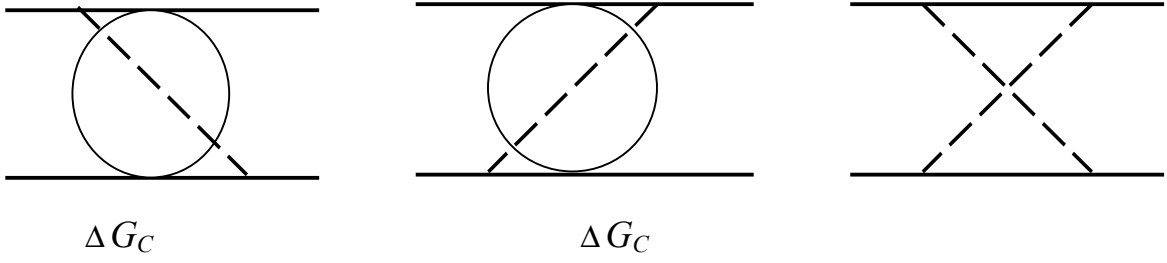

Fig.1.

In the high-frequency domain the Green's function $G_C$ can be expanded in a series and in the investigation of two-photon exchanges, the quasipotential corresponds to the diagrams in *Fig.2*.

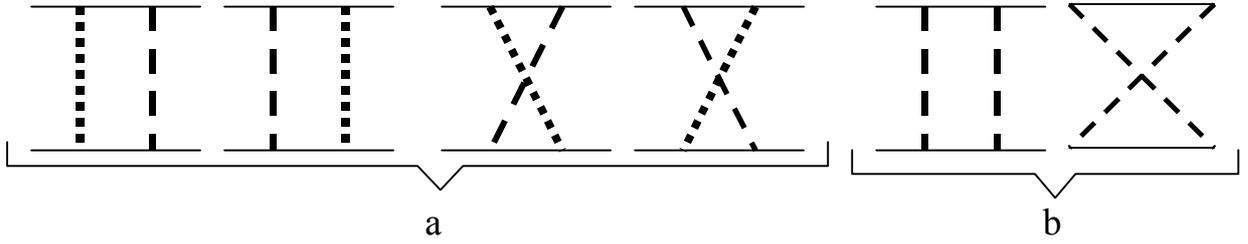

Fig.2.

The results of the calculations, namely, the logarithmic corrections in parameter $\beta$ from diagrams of *Fig.2* are given in *Table 2*.

| diagram | $\Delta E_T$ | $\Delta E_{CT}$ | $\Delta E_{it}$ | $\Delta E_{TT}$ |
|---|---|---|---|---|
| $\dfrac{\alpha^5 \mu^3}{m_1 m_2} \beta^2 \ln \beta^{-1}$ | 0 | 0 | 0 | $\dfrac{2}{\pi}$ |
| $\dfrac{\alpha^6 \mu^3}{m_1 m_2} \ln \beta^{-1}$ | $\dfrac{2\sqrt{2}}{\pi^2}(\ln(1+\sqrt{2})-\sqrt{2})$ | 0 | $-\dfrac{4\sqrt{2}}{\pi^2}(\ln(1+\sqrt{2})-\sqrt{2})$ | 0 |
| $\dfrac{\alpha^6 \mu^3}{m_1 m_2} \beta \ln^2 \beta^{-1}$ | $-\dfrac{1}{\pi^2}$ | $\dfrac{\sqrt{2}}{\pi^2}$ | $\dfrac{2}{\pi^2}$ | 0 |

Table 2.

Note that the result of calculating the logarithmic contribution in $\beta$ of the exchange diagram with two transverse photons obtained in the quasipotential approach coincides with the corresponding quantity in [*2*]. This is the only such correction that must be found in the δ-approximation of the wave function. Other corrections of this type vanish in this approximation and can be obtained only with the exact expressions of the Coulomb wave functions and quasipotential taken into account.

So, these new contributions to the *1S – 2S* shift of the energy levels are of order 10,26 kHz for hydrogen and 66,67 kHz for muonium. The experimentally determined shift of *1S – 2S* levels difference is given by

$$\delta v_{1S\text{-}2S}^{exp} = 2\ 466\ 061\ 413\ 187{,}34\ (84)\ \text{kHz} \tag{50}$$

for hydrogen [*21*] and

$$\delta v_{1S\text{-}2S}^{exp} = 2\ 455\ 528\ 941{,}0\ (9{,}8)\ \text{MHz} \tag{51}$$

for muonium [*22*]. Therefore, the analytic difference of the contributions to the *1S* and *2S* levels that we have found does not exceed the experimental errors.

According to [*21*] new data of the classical Lamb shift of *2S$_{1/2}$ – 2P$_{1/2}$* levels are

$$\Delta E_L^{th} = 1\ 057\ 833\ (4)\ \text{kHz}. \tag{52}$$

$$\Delta E_L^{exp} = 1\ 057\ 845\ (3)\ \text{kHz}. \tag{53}$$

We note that the numerical estimate obtained in our papers for the Lamb shift of 2S energy level in the hydrogen atom makes it possible to increase its theoretical value (52) by 1,5 kHz and to reduce the distinction between the theoretical and experimental value of the classical Lamb shift.

## *REFERENCES*

[1] *M.I. Eides, H. Grotch, V.A. Shelyuto. Phys. Rept. 2001, v.342, p.63.*

[2] *T. Fulton, P.C. Martin, Phys. Rev. 1954, v.95, p. 811.*

[3] *N.A. Boikova, Yu.N. Tyukhtyaev, R.N. Faustov Physics of Atomic Nuclear, Vol.61, no.5, 1998, pp. 781-784.*

[4] *N.A. Boikova, Yu.N. Tyukhtyaev, R.N. Faustov. Physics of Atomic Nuclear, Vol.64, no.5, 2001, pp. 917-920.*

[5] *N.A. Boikova, N.E. Nunko, Yu.N. Tyukhtyaev, R.N. Faustov. Teor.Mat.Phys. 132(3): 1179-1188 (2002).*

[6] *N.A. Boikova, S.V. Kleshchevskaya, Yu.N. Tyukhtyaev, R.N. Faustov. Physics of Atomic Nuclear, Vol.66, no.5, 2003, pp. 893-901.*

[7] *Yu.N. Tyukhtyaev. Teor.Mat.Phys. V.36: p.264 (1978).*

[8] *N.A. Boikova, Yu.N. Tyukhtyaev, R.N. Faustov. Problems of High Energy Physics and Quantum Field Theory.IHEP, Protvino 1983, V.1. p.116.*

[9] *N.E. Nunko, Yu.N. Tyukhtyaev, R.N. Faustov. Yad.Fiz. 1979, V.30, no.2. p.457.*

[10] *N.A. Boikova, N.E. Nunko, Yu.N. Tyukhtyaev, R.N. Faustov. Communications of the JINR. Dubna, 1981, P2-81-582.*

[11] *R.Karplus, A.Klein, Phys.Rev.1952. V.87, N5. P.848.*

[12] *R.N. Faustov. Communications of the JINR. Dubna, 1964, P2-1911.*

[13] *N.E. Nunko, Yu.N. Tyukhtyaev, R.N. Faustov. Communications of the JINR. Dubna, 1973, P2-6996.*

[14] *N.E. Nunko, Yu.N. Tyukhtyaev. Teor.Mat.Fiz. 1972. V.12, no.1. p.56.*

[15] *Yu.N. Tyukhtyaev, R.N. Faustov. Communications of the JINR. Dubna, 1986, P2-86-281.*

[16] *N.E. Nunko, Yu.N. Tyukhtyaev, R.N. Faustov. Problems of High Energy Physics and Quantum Field Theory. IHEP, Protvino 1983, V.1. p.104.*

[17] *N.A. Levchenko, Yu.N. Tyukhtyaev, R.N. Faustov. Yad.Fiz. 1980, V.32, no.6 (12). p. 1656.*

[18] *N.A. Levchenko, N.E. Nunko, Yu.N. Tyukhtyaev. Communications of the JINR. Dubna, 1979, P2-12355.*

[19] *G.P.Lepage. Phys. Rev. 1977. V. A16, N.3. P.863.*

[20] *G.T. Bodwin, D.R. Yennie. Phys. Rept. 1978. V. C43, N.6. P. 267.*

[21] *M.I. Eides, H. Grotch, V.A. Shelyuto. Hepph / 0002158.*

[22] *V. Meyer, S.N. Bagaev, P.E. Baird, et al. Phys. Rev. 2000. A51. P. 1854.*